# Cryogenic Nano-Imaging of Excitons in a Monolayer Semiconductor

*Anna S. Roche[1], Michael R. Koehler[2], David G. Mandrus[3-5], Takashi Taniguchi[6], Kenji Watanabe[7], John R. Schaibley[1], Brian J. LeRoy[1]*

[1]Department of Physics, University of Arizona, Tucson, Arizona 85721, USA

[2]IAMM Diffraction Facility, Institute for Advanced Materials and Manufacturing, University of Tennessee, Knoxville, TN 37920

[3]Department of Materials Science and Engineering, University of Tennessee, Knoxville, Tennessee 37996, USA

[4]Materials Science and Technology Division, Oak Ridge National Laboratory, Oak Ridge, Tennessee 37831, USA

[5]Department of Physics and Astronomy, University of Tennessee, Knoxville, Tennessee 37996, USA

[6]Research Center for Materials Nanoarchitectonics, National Institute for Materials Science, 1-1 Namiki, Tsukuba 305-0044, Japan

[7]Research Center for Electronic and Optical Materials, National Institute for Materials Science, 1-1 Namiki, Tsukuba 305-0044, Japan

ABSTRACT

Optical measurements of 2D semiconductors have primarily relied on far-field spectroscopy techniques which are diffraction limited to several hundred nanometers. To precisely image nanoscale spatial disorder requires an order of magnitude increase in resolution capabilities. Here, we present a spatially resolved study of the exciton spectra of monolayer $MoSe_2$ in the visible range using a cryogenic scattering-type scanning near-field optical microscope (s-SNOM) operating down to 11 K. Mapping the exciton resonance across an hBN encapsulated $MoSe_2$ monolayer, we achieve sub-50 nm spatial resolution and energy resolution below 1 meV. We further investigate the material's near-field spectra and dielectric function, demonstrating the ability of cryogenic visible s-SNOM to reveal nanoscale disorder. Comparison to room temperature measurements illustrate the enhanced capabilities of cryogenic s-SNOM to reveal fine-scale material heterogeneity. These results establish cryogenic visible s-SNOM as an effective nanoscale excitonic probe offering valuable insights into 2D material heterogeneity and nanoscale sensing.

TEXT

Atomically thin monolayer transition metal dichalcogenide (TMDs) semiconductors are a promising platform for fundamental studies of two-dimensional (2D) systems with wide reaching applications in optoelectronics[1,2]. A direct consequence of their reduced dimensionality is the formation of strongly bound electron-hole pairs, or excitons, which dominate the material's optical properties across visible and near-infrared frequencies due to their large binding energy and prominent oscillator strength[3,4,5,6,7]. These excitons are highly sensitive to any extrinsic disorder[8,9] and to accurately investigate and understand the optical effects of inhomogeneity in such materials requires nanoscale probes with resolution capabilities beyond those of far field optics[10]. Previous studies of TMD materials with resolution below the diffraction limit have employed scanning probe techniques such as tip enhanced photoluminescence (TEPL)[11,12,13,14] and Raman spectroscopy (TERS)[15,16,17,18] and scattering type scanning near-field optical microscopy (s-SNOM)[19]. While there has been research on the near field optical response at room temperature, a low temperature investigation of 2D semiconductors near field response is lacking. Previous s-SNOM studies on TMDs showed a homogenous spatial signal at room temperature[19]. Cryogenic s-SNOM enables us to resolve sub 1 meV shifts in excitonic energy as the resonance linewidth narrows to 4 meV at 11 K compared to ~30 meV at room temperature. Furthermore, the integration of a resonant narrow linewidth tunable continuous wave laser with a cryogenic s-SNOM opens new opportunities for probing nanoscale optical properties. These advanced capabilities allow for 2D materials and surfaces to be probed at the nanoscale providing important insights towards the future of quantum material engineering in van der Waals materials and nanoscale sensing.

In this work, we image the nanoscale excitonic response of a hexagonal boron nitride (hBN) encapsulated $MoSe_2$ monolayer at cryogenic temperatures with resonant visible (~750 nm) s-SNOM. The nanoscale optical measurements were performed using a cryogenic s-SNOM (cryo-neaSCOPE) which is operational from room temperature to 11 K, customized to incorporate a tunable high resolution continuous wave laser. The experimental technique of s-SNOM[20,21,22,23,24,25] is built upon the working principles of atomic force microscopy (AFM) (Fig.

1A). A metal coated AFM tip is illuminated by a focused laser. The sharp, metallic tip acts like an antenna[26], increasing the coupling of the excitation light to produce an enhanced optical field at the tip apex[27]. The near field optical resolution is limited by the AFM tip apex allowing sub-diffraction optical resolution on the order of 50 nm. The enhanced field confinement is 10 times tighter than the diffraction limit at the low temperature MoSe$_2$ exciton resonance (750 nm). When the tip is brought close to the sample, the near field interaction between the tip and sample modifies the light backscattered from the tip. The back scattered light carrying the nanoscale optical information is collected at a silicon detector in the far field. Pseudo-heterodyne interferometric detection (PHID)[28] relies on demodulating this collected field at higher harmonics of the oscillating AFM tip frequency (Ω) to isolate the near field signal from the otherwise dominant far field background signal. The near field modification from the sample can be described with a complex valued scattering coefficient $\sigma = se^{i\theta}$, as $E_{scat} = \sigma E_0$. Alongside superior background suppression, PHID allows the amplitude and phase of the backscattered near-field signal (s and θ of $\sigma = se^{i\theta}$) to be recorded separately. The near-field amplitude and phase spectra is used to extract local information on the TMD's dielectric function and excitonic resonances. The illuminated tip is raster scanned across the TMD sample to simultaneously measure the topography, near field amplitude, and phase. The AFM topography of the monolayer MoSe$_2$ flake is shown in Figure 1B. Figures 1C and D illustrate the near-field amplitude and phase measured simultaneously at ω= 1.66 eV (~747 nm) and 1.58 eV (~785 nm) at 11K and room temperature respectively. Notably, the observed near field signal across the flake is 1.25 times stronger and the narrow spectral lines allow for the observation of inhomogeneity that is not spectrally resolvable at room temperature. All measurements presented were measured at 11 K.

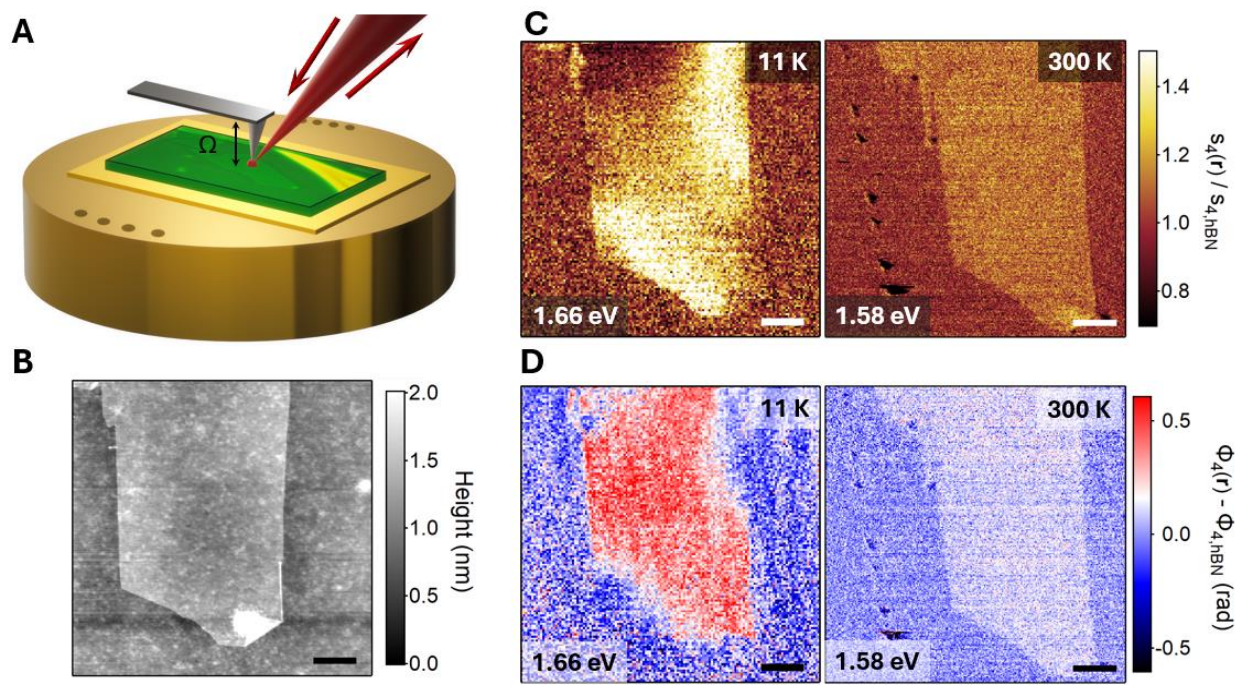

**Figure 1.** Near-field measurements at room and cryogenic temperature. **A** Schematic of the s-SNOM experimental set-up. A tunable visible laser is focused on a metallic AFM tip, operating in tapping mode at frequency Ω, to produce a nanoscale enhanced spot at the tip apex. The near field spot interacts with the sample (depicted in the optical microscope image) below the tip and backscatters back to the focusing optic to be sent to a far field detector. The cryogenic system is installed on a cold head, illustrated by the gold disk, and held in an optical cryostat at 11 K. **B** A cryogenic AFM topography image of the hBN encapsulated $MoSe_2$ sample, measured simultaneously with the near-field images. **C** Normalized near-field amplitude ($s_4$) image of the sample taken at the 4th harmonic of the tapping frequency at 11 K (300 K) on the left (right) shown on the same color scale. The 11 K image was taken at an excitation energy of 1.66 eV and the room temperature image at an energy of 1.58 eV. **D** The corresponding normalized near-field phase ($\Phi_4$) image of the sample. Scale bars are 1 μm.

To investigate the excitonic behavior at low temperature, we spatially mapped the neutral ($X_0$) exciton resonance as a function of excitation energy. We repeated scans over the area of the MoSe$_2$ flake while stepping the excitation energy through the $X_0$ resonance near 1.65 eV. Figures 2 (A-E) display near field amplitude and (F-J) phase images taken with 50 nm spatial resolution as the excitation energy was changed in ~2 meV steps from 1.653 to 1.662 eV. S-SNOM signals are understood in relationship to a control material which in this case is taken to be the hBN substrate. The amplitude and phase at the fourth harmonic of the tip oscillation frequency were normalized against the hBN signal as $s_4(r)/s_{4,hBN}$ and $\theta_4(r) - \theta_{4,hBN}$[21], since hBN has no energy dependent optical response in the energy range of the $X_0$ resonance. The MoSe$_2$ flake can be clearly seen in Fig. 2A as the area in yellow, corresponding to the AFM topography image of the flake geometry in Fig. 1B. The expected excitonic response is observed in this collection of images as the signal increases to a larger value than the surrounding bare hBN before becoming smaller than it as the energy increases. To understand the information depicted by the s-SNOM images, we focus on the area within the red box in Fig. 2A. The amplitude intensity is relatively high at 1.653 eV, then as the energy increases, the intensity peaks at 1.655 eV and quickly drops above this energy, illustrating the resonance behavior. Looking at the same region in the phase images, we see the intensity is initially low, reaches a maximum around 1.655 eV, and falls back down, illustrating a peak in the phase centered around the resonance at 1.655 eV. This demonstrates how we can track the near field intensity and phase to observe the expected resonance behavior, corresponding to the $X_0$ MoSe$_2$ exciton.

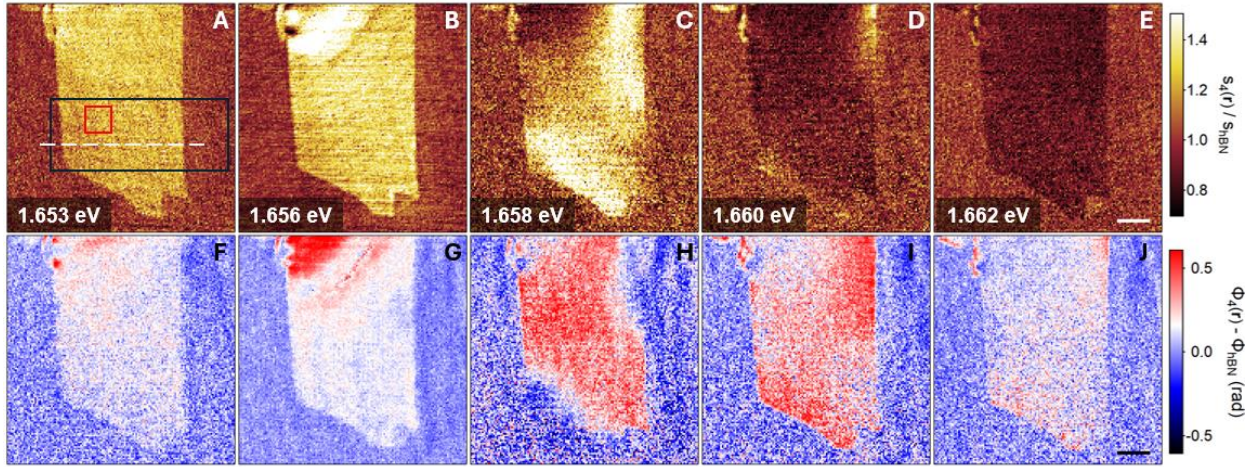

**Figure 2.** Near field images under varying excitation energy. **A-E** Normalized amplitude $s_4$ image of the sample taken at excitation energies 1.653, 1.656, 1.658, 1.660, 1.662 eV. The red square demonstrates a smaller region of interest for visualizing resonance behavior. The white dashed line corresponds to the line cuts in Figs. 3A and B. The black rectangle corresponds to the region mapped in Fig. 4. **F-J** Corresponding normalized phase $\Phi_4$ images measured at the same energy shown in the above amplitude scans. All images measured at the 4th harmonic. Scale bars are 1 μm.

Now, looking at the entire flake; at low temperature, as the exciton resonance narrows, greater spatial disorder of the $X_0$ resonance is observed. Markedly, such disorder is not observed at room temperature on the same sample (Figure S1) and in previous works[19,29]. Figure 2 shows that there is resolvable spatial disorder of the $X_0$ resonance throughout the TMD flake, demonstrated by the variation in the amplitude and phase signal intensity as the excitation energy is swept.

To quantify the homogeneity of the sample, we measured the low temperature excitonic spectra across the TMD sample while simultaneously extracting the spatial and energy resolution capabilities of our experimental setup. Line scans with 30 nm spatial resolution were taken horizontally across the sample, along the dashed line in Fig. 2A, as the excitation energy was swept from 1.651 to 1.669 eV in 0.22 meV steps. The near field amplitude (Fig. 3A) and phase (Fig. 3B) of the MoSe$_2$ are displayed as a function of excitation energy and distance along this line. A steep redshift in the central energy of the $X_0$ resonance is observed within 100 nm of the left edge, attributable to the change in dielectric function, followed by a gradual redshift from 1.662 to 1.660 eV as the sample is scanned across to the right edge. A similar flake edge energy shift of 4 meV over ~30-80 nm has been observed in tip-enhanced photoluminescence images of physical vapor transport grown WSe$_2$[12]. Figures 3C and 3D display vertical line cuts through Figs. 3A and 3B along the blue dashed lines. These line cuts, construct a near field spectrum of MoSe$_2$ at 11 K. The amplitude and phase spectra follow the expected resonance behavior. The amplitude spectrum shows an S-curve, centered around the same energy as the single peak in the phase spectrum. These spectra can be used to extract the complex dielectric function of the TMD monolayer when fit using the well-established Finite Dipole Model (FDM)[30,31]. We modelled the measured near field response with the FDM, assuming a Lorentz oscillator form of the complex permittivity[19,32,33] near resonance $\varepsilon(\omega) = \varepsilon_1(\omega) + i\varepsilon_2(\omega)$:

$$\varepsilon(\omega) = \varepsilon_\infty - \frac{\hbar c}{\omega_0 d} \frac{\gamma_r}{(\omega_0 - \omega) + i\left(\frac{\gamma_{nr}}{2} + \gamma_d\right)}$$

where $\varepsilon_\infty$ is the high frequency permittivity of the sample, $\omega_0$ is the center energy of the oscillator, $\gamma_r$ is the radiative decay rate, $\frac{\gamma_{nr}}{2} + \gamma_d$ are the non-radiative and dephasing decay rates, and d is the monolayer thickness. The amplitude of the near field signal is related to the real part of the dielectric function while the phase is related to the imaginary part. The results of the FDM model are shown as solid lines in Figs. 3C and 3D alongside the modeled dielectric function in Fig. 3E. The dielectric response matches the observed near field signal and previous far-field work investigating ultra-clean hBN encapsulated TMDs[32,34]. The calculated dielectric function has a zero crossing in $\varepsilon_1$, only seen in samples with a narrow resonance linewidth. A negative $\varepsilon_1$ is a known signature for 2D exciton-polariton formation in ultra-clean monolayer TMDs. The ratio $-\varepsilon_1/\varepsilon_2$ is displayed in the Fig. 3D inset, illustrating the optimal energy range to observe 2D exciton polaritons[34]. The ability to measure a negative permittivity on the nanometer scale paves the way for the direct imaging of 2D exciton-polariton behavior.

From the high-resolution results in Fig. 3A we can extract the spatial resolution of the near-field signal. To extract the spatial resolution of our system, a horizontal line cut across the TMD edge in Fig. 3A is displayed in Figure S2 The near field amplitude intensity as the AFM tip steps from hBN to MoSe$_2$ was fit with a corresponding line spread function from which we measure a spatial resolution of 45 nm, an order of magnitude increase in resolution from diffraction limited systems. From the variation of the amplitude and phase signals with position, we can clearly

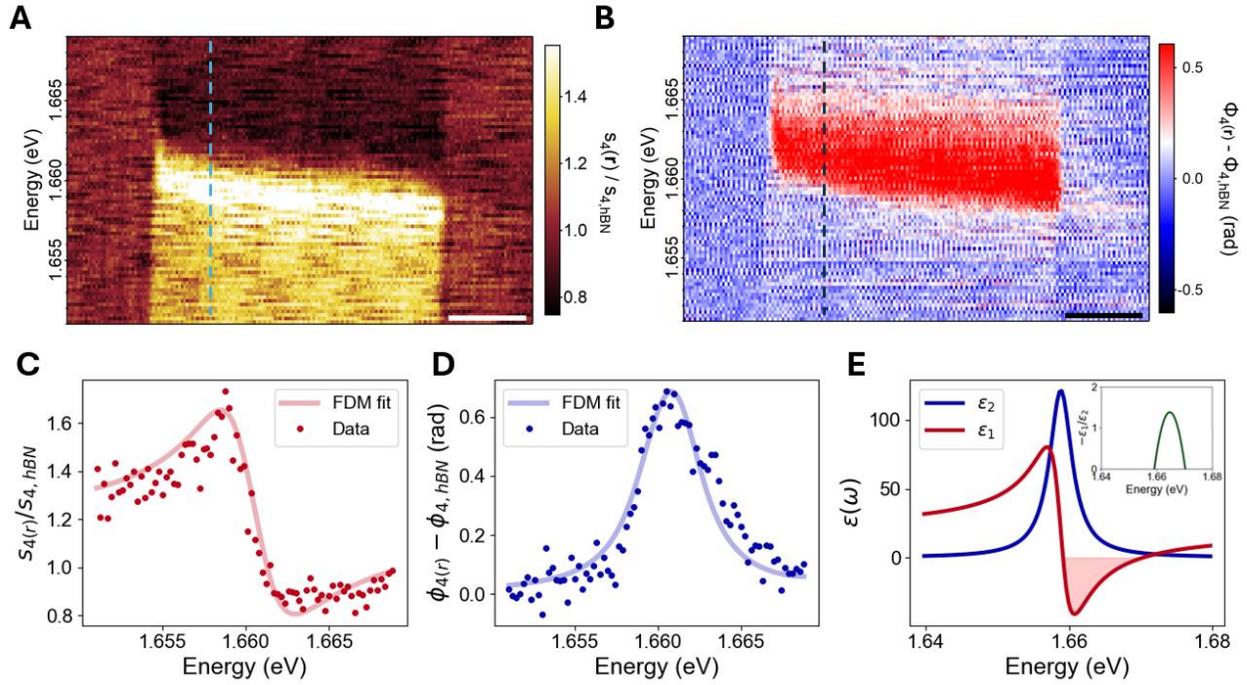

**Figure 3**. Extraction of dielectric function. **A,B** Variation of the near-field amplitude $s_4$ and phase $\Phi_4$ signal across the line scan shown in Fig. 2A as a function of excitation energy (left axis). Scale bars are 1 μm. **C,D** Near-field (C) amplitude and (D) phase spectra extracted from 3A and 3B along the dashed blue line. The spectra are fitted with the finite dipole model (FDM) illustrated by the solid curves through the data. **E** Calculated complex dielectric function of monolayer hBN encapsulated $MoSe_2$ at 11 K extracted from the FDM fits shown in C,D. Inset: Plotted $-\varepsilon_1/\varepsilon_2$ illustrating the energy region to explore in future 2D exciton polariton studies

observe changes in the center energy of the $X_0$ resonance of less than 1 meV. The measurements in Figs. 3C and 3D were repeated at increasing temperatures and show the expected red shift of the $X_0$ resonance and its spectral broadening (Figure S3).

To continue our high-resolution characterization, we mapped the near field spectra at every pixel across a 2 by 5.5 micron region of the TMD flake, depicted by the black box in Fig. 2A. 2D scans were repeated across the flake from 1.651 to 1.668 eV in 0.65 meV steps with a spatial resolution of 30 nm per pixel. Taking the amplitude and phase signal as a function of excitation energy we measure the near field spectra (like in Figs. 3C and 3D) at each pixel. Following the same FDM fitting procedure, we calculate the complex dielectric function from each near field spectra. Figure 4A shows the extracted $X_0$ center energy, $\omega_0$, across the MoSe$_2$ sample. The center energy of the $X_0$ resonance shows a clear spatial variation throughout the flake, corresponding to the redshift behavior observed in Fig. 3A. A far field PL map of the MoSe$_2$ flake (Figure S5 A) shows a roughly uniform center energy of the $X_0$ peak, demonstrating the observed disorder to be only visible due to the increased resolution of near-field techniques. The MoSe$_2$ PL spectra showed a clearly resolvable trion peak (Figure S5C) throughout the map; the ratio of the neutral and trion peak intensities showed a roughly constant sample doping across the region depicted in Fig. 4 (Figure S5 B). Fig. 4B displays a histogram of the center energy values across the region shown in the corresponding image. The clear asymmetric distribution stems from the shift in $\omega_0$ from 1.662 eV to below 1.659 eV moving across the flake from the bottom left corner. These fine changes in energy at sub 50 nm resolution are only observable in s-SNOM measurements while a far-field

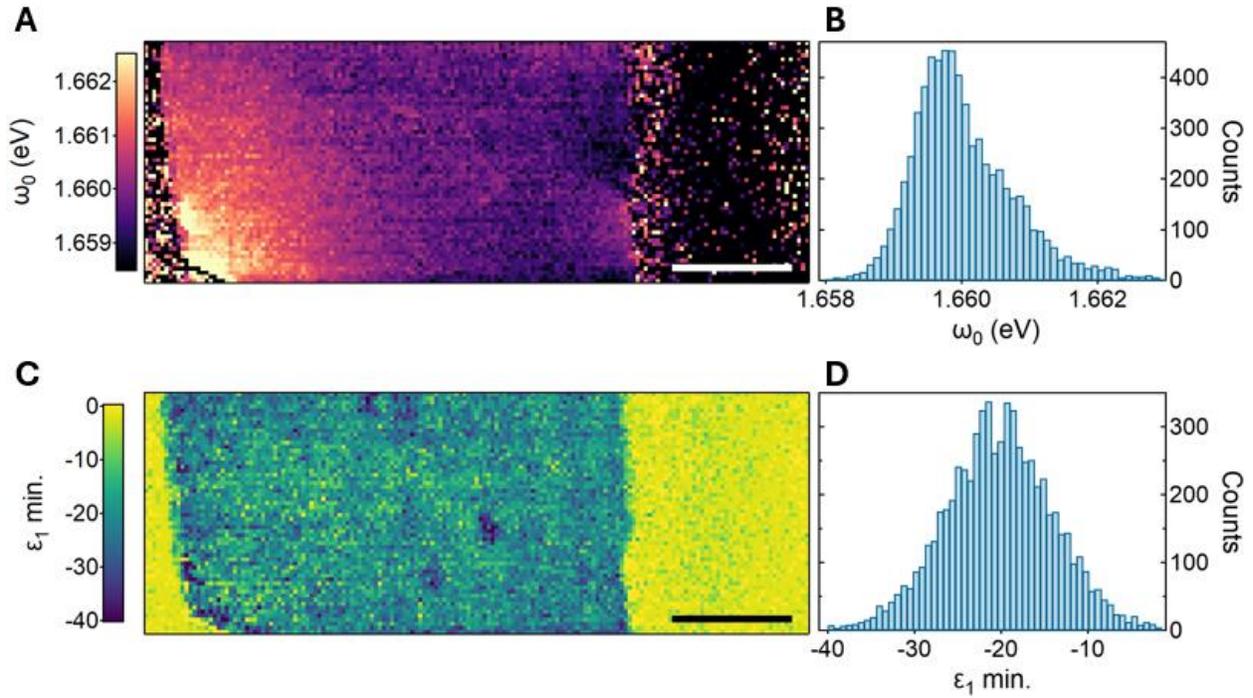

**Figure 4**. Spatial variation of resonance and dielectric function. **A** $X_0$ resonance energy $\omega_0$ value across the sample region shown in the black rectangle in Fig. 2A extracted from the complex dielectric function at each pixel. Spatial disorder in the $X_0$ resonance energy is observed across the MoSe$_2$ flake. **B** Histogram of the resonance energy values across the TMD region in A. **C** Minimum value of $\varepsilon_1$ across the same sample region. **D** Corresponding histogram of $\varepsilon_1$ minimum across the TMD region. Scale bars are 1 μm

measurement only observes the broader distribution shown in Fig. 4B. The decay parameters and high-frequency permittivity, $\varepsilon_\infty$, show less spatial dependence and are roughly constant over the area investigated (Figure S6). The calculated dielectric function shows a negative valued $\varepsilon_1$ across the entirety of the flake for certain energies demonstrating the narrow linewidth of the exciton resonance. Figure 4C shows the minimum value $\varepsilon_1$ reaches at each point throughout the flake along with its distribution in Fig. 4D indicating uniform narrow linewidths.

This work confirms cryogenic visible s-SNOM as a promising measurement tool, providing previously unattainable nanoscale optical information of the local dielectric function of a TMD monolayer. We simultaneously identified the energy and spatial resolution capabilities of cryogenic visible s-SNOM by extracting a high-resolution spatial map of the $X_0$ resonance across a TMD flake. The high spatial resolution allows energy resolved measurements at sub meV resolution which are not affected by spatial disorder in the TMD as is typical in a far-field measurement. With our sub 50 nm resolution capabilities we observe spatial variation undetectable by traditional far-field optical techniques, necessitating further studies into the origin of the observed sample disorder. From this low-temperature ultra-clean, hBN encapsulated $MoSe_2$, we confirm previous reports of narrow linewidth samples giving rise to negative $\varepsilon_1$, motivating additional investigations into directly imaging 2D exciton-polaritons in TMDs. Cryogenic nanoimaging also opens the door to using 2D material excitons as precise sensors of local strain, disorder, and doping.

## ASSOCIATED CONTENT

**Supporting Information**

The following files are available free of charge.

Supporting Information For: Cryogenic Nano-Imaging of Excitons in a Monolayer Semiconductor

Details on sample fabrication, optical measurements, and finite dipole modeling. Additional figures and details on room temperature s-SNOM measurements, temperature dependence of the near field spectra, extraction of the optical spatial resolution, and far field photoluminescence maps. (PDF)

AUTHOR INFORMATION

**Corresponding Author**

Brian LeRoy - Department of Physics, University of Arizona, Tucson, Arizona 85721, USA.

leroy@arizona.edu

**Author Contributions**

AR, JRS, and BJL conceived the project. JRS and BJL supervised the project. AR performed the experiments. AR and BJL analyzed the data with input from JRS. MRK and DGM provided and characterized the bulk $MoSe_2$ crystals. TT and KW provided bulk hBN crystals. AR, JRS, and BJL wrote the paper. All authors discussed the results.

**Funding Sources**

JRS and BJL acknowledge support from Air Force Office of Scientific Research Grant Nos. FA9550-22-1-0312 and FA9550-22-1-0220. BJL acknowledges support from the National Science Foundation Grant No. ECCS-2122462. JRS acknowledges support from Air Force Office of Scientific Research Grant Nos. FA9550-20-1-0217 and FA9550-21-1-0219. DGM acknowledges support from the Gordon and Betty Moore Foundation's EPiQS Initiative, Grant GBMF9069. KW and TT acknowledge support from the JSPS KAKENHI (Grant Numbers 21H05233 and 23H02052), the CREST (JPMJCR24A5), JST and World Premier International Research Center Initiative (WPI), MEXT, Japan.

ACKNOWLEDGMENT

We thank Rolf Binder and Vasili Perebeinos for helpful discussions.

ABBREVIATIONS

s-SNOM, scattering-type scanning near field optical microscopy; TMD, transition metal dichalcogenide; 2D, two-dimensional; $MoSe_2$, molybdenum diselenide; $WSe_2$, tungsten diselenide; hBN, hexagonal boron nitride; AFM, atomic force microscopy; $X_0$, neutral exciton; FDM, finite dipole moment; PL, photoluminescence.

# Supporting Information For: Cryogenic Nano-Imaging of Excitons in a Monolayer Semiconductor

Anna S. Roche[1], Michael R. Koehler[2], David G. Mandrus[3-5], Takashi Taniguchi[6], Kenji Watanabe[7], John R. Schaibley[1], Brian J. LeRoy[1]

[1]Department of Physics, University of Arizona, Tucson, Arizona 85721, USA

[2]IAMM Diffraction Facility, Institute for Advanced Materials and Manufacturing, University of Tennessee, Knoxville, TN 37920

[3]Department of Materials Science and Engineering, University of Tennessee, Knoxville, Tennessee 37996, USA

[4]Materials Science and Technology Division, Oak Ridge National Laboratory, Oak Ridge, Tennessee 37831, USA

[5]Department of Physics and Astronomy, University of Tennessee, Knoxville, Tennessee 37996, USA

[6]Research Center for Materials Nanoarchitectonics, National Institute for Materials Science, 1-1 Namiki, Tsukuba 305-0044, Japan

[7]Research Center for Electronic and Optical Materials, National Institute for Materials Science, 1-1 Namiki, Tsukuba 305-0044, Japan

**Table of Contents**

1. **Sample Fabrication**

2. **Optical Measurements**

3. **Room Temperature Measurements**

4. **Spatial Resolution**

5. **Temperature Dependent Near-field Spectra**

6. **Far field Photoluminescence**

7. **Finite Dipole Modeling**

1. **Sample Fabrication**

Atomically thin layers of hBN and MoSe$_2$ were isolated using standard mechanical exfoliation techniques, then stacked into an hBN encapsulated MoSe$_2$ monolayer heterostructure using a polymer based dry-transfer method[1]. The top (bottom) hBN was 2 nm (23 nm) thick. To ensure a clean surface suitable for scanning probe measurements the heterostructure was picked up using a polypropylene carbonate (PPC) coated polydimethylsiloxane (PDMS) stamp in reverse order and flipped onto a clean SiO$_2$/Si wafer. The sample was then annealed in vacuum at 275 degrees C for 6 hours and AFM cleaned[2] prior to s-SNOM measurements.

2. **Optical Measurements**

All optical measurements were performed in a cryo-neaSCOPE using a Pt/Ir-coated silicon AFM tip. The cryogenic s-SNOM is held in an optical cryostat at 11 K. Near field amplitude and phase images were measured with a tunable visible Ti-S laser (MSquared SolsTiS), fiber coupled into the s-SNOM optical system and subsequently power (4 mW) and mode locked to ensure proper stability. All near field images were detected at the 4th harmonic of the AFM tapping frequency. Far field PL maps were measured with a 633 nm diode laser.

3. **Room Temperature Measurements**

Near-field images of the MoSe$_2$ device confirmed that the spatial disorder is only observable at low temperatures, where the exciton resonance is sufficiently narrow to observe small variations in the center energy of the X$_0$ resonance. At room temperature the exciton resonance width is ~40 meV[5], roughly an order of magnitude larger than that at 11 K, obscuring the sub 2 meV spatial variation present throughout the sample. A constant response in both amplitude and phase across the TMD flake is observed in Fig. S2 as the excitation energy is varied

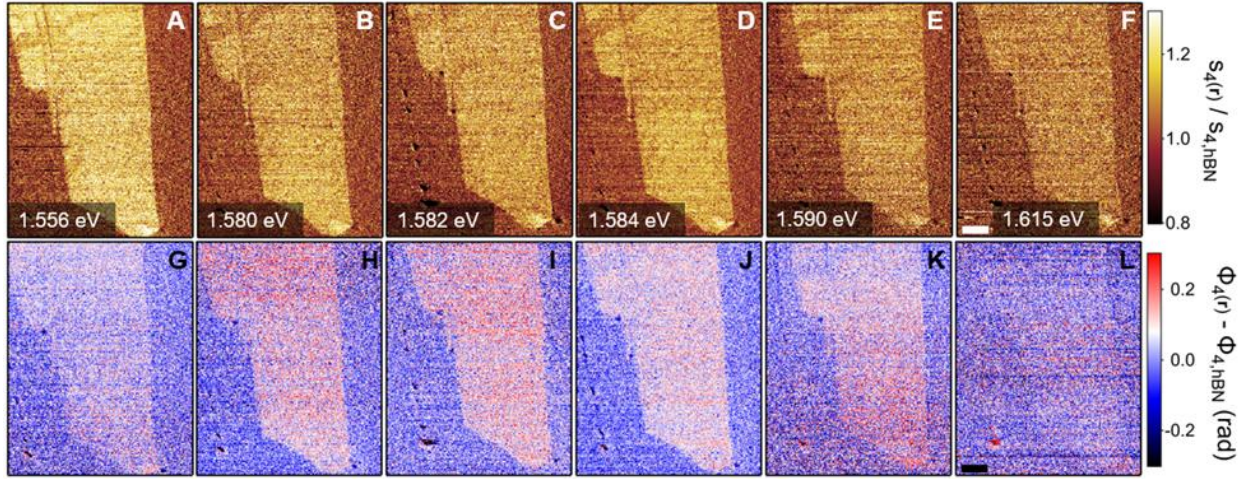

**Figure S2:** Room temperature comparison of near field images under varying excitation energy. **A-F** Normalized amplitude $s_4$ images of the sample taken at excitation energies of 1.556, 1.580, 1.582, 1.584, 1.590, 1.615 eV. **G-L** Corresponding normalized phase $\Phi_4$ images measured at the same energy shown in the above amplitude scans. All images measured at the 4th harmonic. Scale bars are 1 μm.

4. **Spatial Resolution**

The spatial resolution limit of s-SNOM depends on the AFM tip radius, which range from roughly 25-100 nm, as well as the harmonic *n* measured, the latter referred to as effective tip sharpening[6]. To extract this limit, we performed line scans across the MoSe$_2$ edge with 10 nm/pixel resolution. The line cuts were taken at 1.65 eV which is above the $X_0$ resonance, resulting in a uniform signal across the flake to best fit the edge step. From the near-field amplitude intensity variation from the hBN to TMD step we extract a spatial resolution of 45 nm from our line spread function, as calculated as the derivative of the edge step function.

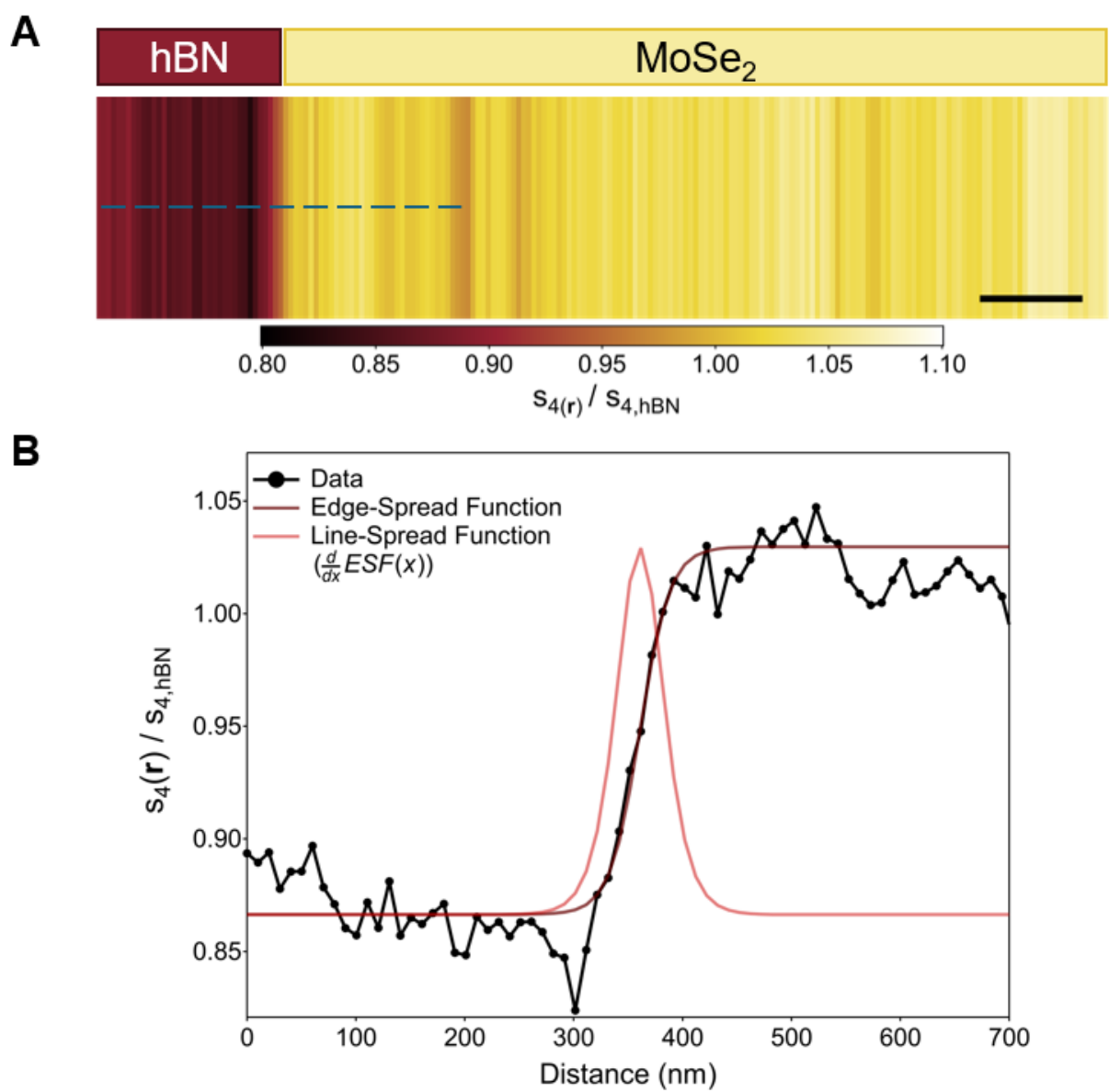

**Figure S2.** Spatial resolution of near-field signal. **A** Normalized amplitude $s_4$ image of the sample taken across the MoSe$_2$ monolayer edge with 10 nm resolution at energy 1.65 eV. Scale bar is 200 nm. **B** Line cut across the edge taken at the blue dashed line. This data set represents our edge-spread function (ESF). The line-spread function (LSF) is calculated as the spatial derivative of the ESF, and shown in light red. The resolution limit is given by the FWHM of the LSF, to be 45 nm.

## 5. Temperature Dependent Near-field Spectra

The exciton resonance width and center energy depend on the sample temperature[5], demonstrating a well-established broadening and redshift as the sample temperature increases. We observe the expected broadening and center energy shifts from increasing temperature.

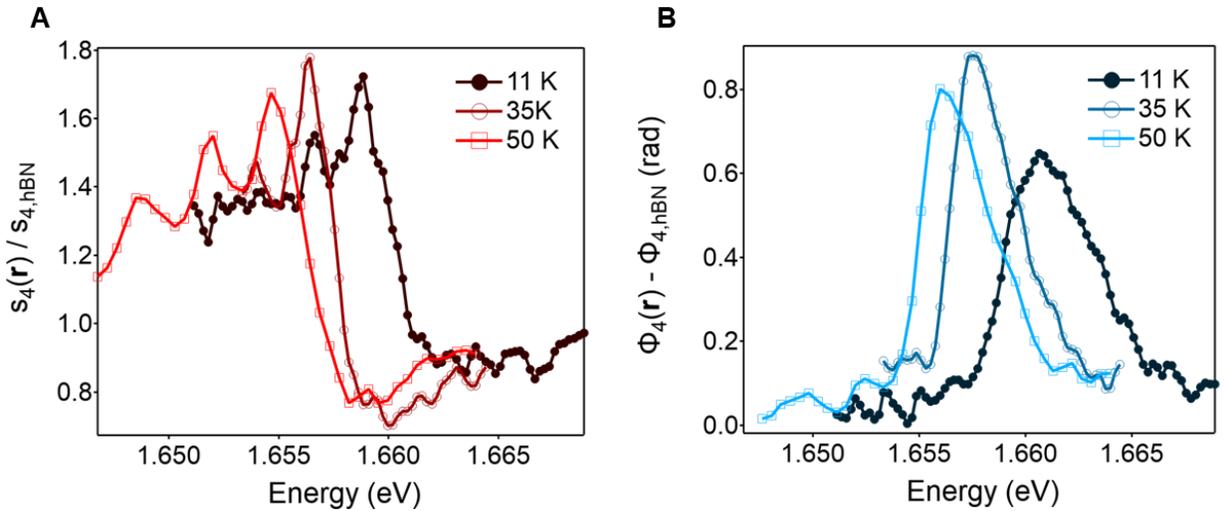

**Figure S3.** Temperature dependent near field spectra. The measurements in Fig. 2 were repeated at 35 K and 50 K to confirm the expected resonance widening and center energy redshift with increasing temperature. **A** The near-field amplitude spectra extracted from the measured variation of the near-field amplitude $s_4$ averaged across a line scan like that shown in Fig. 2A as a function of excitation energy at 11 K, 35 K, and 50 K. **B** Near-field phase spectra corresponding with those shown in A, at 11 K, 35 K, and 50 K.

## 6. Far field Photoluminescence

Traditional far-field measurements have not resolved the spatial disorder on the $X_0$ resonance uncovered in near-field measurements. To confirm this within our investigated device, we performed far-field photoluminescence (PL) maps of the $MoSe_2$ sample. PL maps were taken within the s-SNOM set up using a 633 nm diode laser. The laser was focused on the sample using a parabolic mirror with a 0.39 NA. The PL was recollected at the parabolic mirror and sent through a long pass filter to block the laser. The PL was magnified and imaged with a cooled CCD (Andor Newton).

Clear neutral exciton ($X_0$) and charged trion (T) peaks were resolvable in our device as shown in Fig. S5 indicative of a doped sample[7]. In addition to extracting a spatial map of the center energy of the $X_0$ PL peak, we plotted the the ratio of the T to $X_0$ PL intensity as a map of the doping across the $MoSe_2$ flake. The region imaged in Fig. 4, is shown by the black box in Fig. S6. The $X_0$ PL center energy shows the spatial disorder resolved in Fig. 4 is obsreved only in near-field measurements. The map of the trion to $X_0$ PL intensity shows the region measured in Fig. 4 to have roughly constant doping.

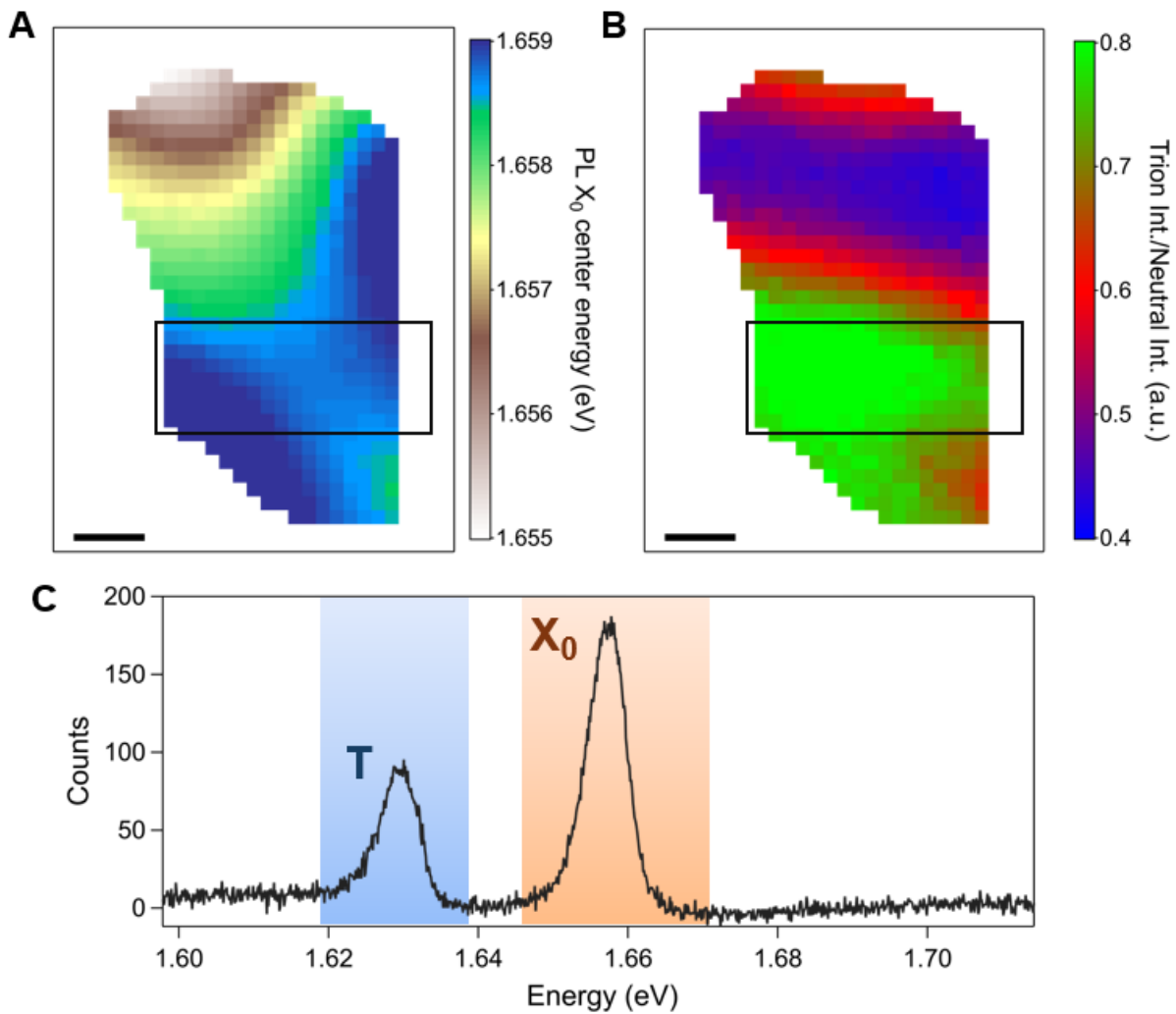

**Figure S5.** Far Field photoluminescence. **A** Photoluminescence map of the investigated MoSe$_2$ flake. Color bar matches the energy scale in Fig.4A to demonstrate the decrease in spatial and energy resolution using traditional far field techniques. The black box approximately outlines the region measured in Fig. 4A. **B** The ratio of the T peak intensity to the $X_0$ peak is used to investigate if the disorder seen in Fig. 4A might stem from sample doping. The sample doping is shown to be roughly constant throughout the region measured in Fig. 4A, suggesting the $X_0$ resonance energy disorder originates from a different source. Scale bar is 5 microns. **C** Far field photoluminescence spectrum of the hBN encapsulated MoSe$_2$ measured in all near-field images.

## 7. Finite Dipole Model Methods

Finite Dipole Model (FDM) fitting was performed using the open-source python package snompy[3]. We modeled our device as a multi-layer stack, following the transfer matrix method[4]. From top to bottom the stacks and thickness are as follows; air (infinite environment), hBN (2 nm), MoSe$_2$ (0.7 nm), hBN (23 nm), SiO$_2$ (infinite substrate). The chosen dielectric functions of hBN and SiO$_2$ in the visible range were 3.8 and 2.1 respectively. The ellipsoidal tip model had a length of 350 nm and a radius of curvature of 25 nm, chosen to match the geometry of the AFM probe used for the measurements in this work. The Lorentz oscillator model of the complex dielectric response of MoSe$_2$ was calculated through a recursive fitting process with our near field spectra data and the FDM simulation results. From the complex Lorentz oscillator equation

$$\varepsilon(\omega) = \varepsilon_\infty - \frac{\hbar c}{\omega_0 d} \frac{\gamma_r}{(\omega_0 - \omega) + i\left(\frac{\gamma_{nr}}{2} + \gamma_d\right)}$$

the high frequency permittivity of the sample ($\varepsilon_\infty$), the center energy ($\omega_\infty$) of the oscillator, and the radiative ($\gamma_r$), non-radiative and dephasing decay rates $\left(\frac{\gamma_{nr}}{2} + \gamma_d\right)$ were the input parameters of the recursive fit. These values were varied to minimize the difference between the calculated near field response and measured data.

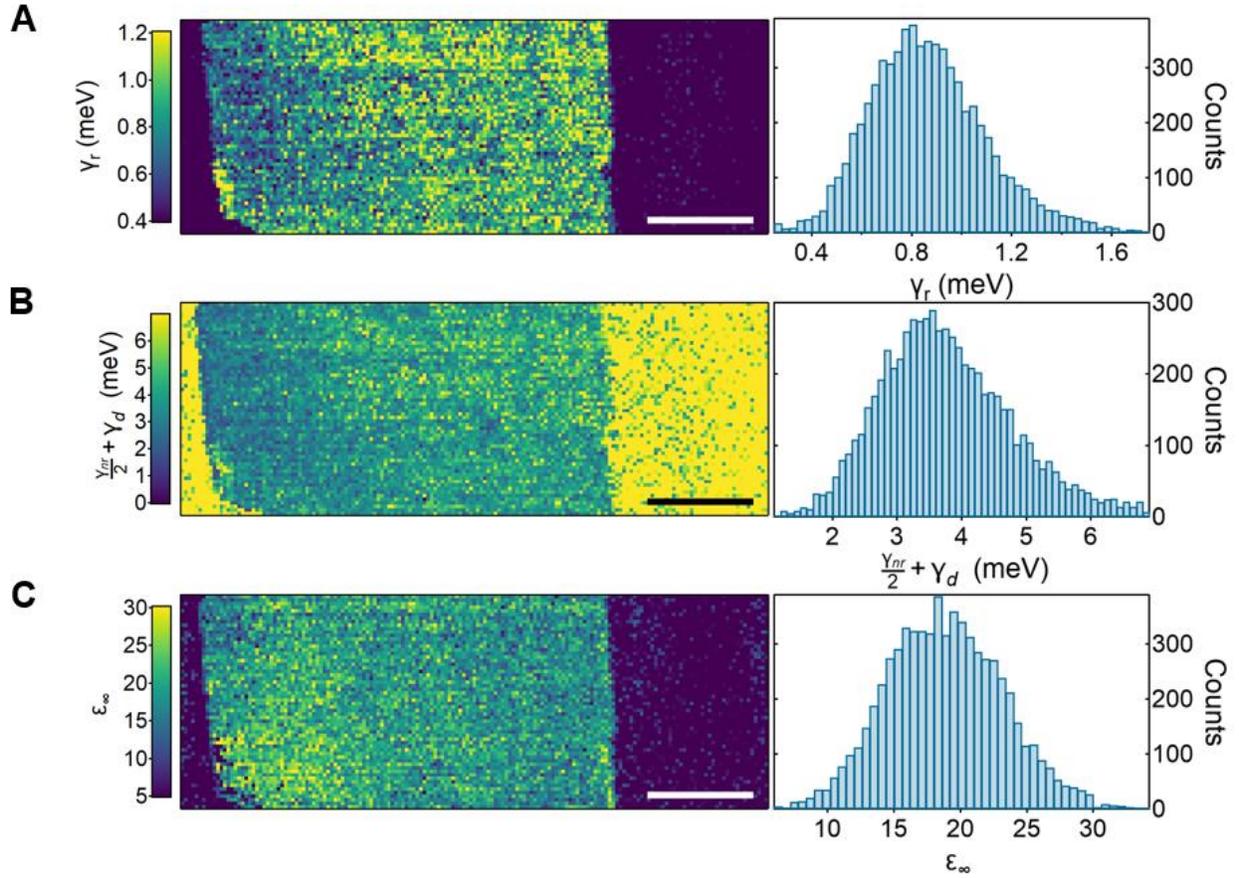

**Figure S6.** FDM extracted complex dielectric function parameter fit results. **A** The radiative decay rate, $\gamma_r$, value across the sample region shown in the black rectangle in Fig. 2A extracted from the complex dielectric function at each pixel **B** The non-radiative and dephasing decay rate $\frac{\gamma_{nr}}{2} + \gamma_d$ value across the sample region. **C** The high frequency permittivity $\varepsilon_\infty$ values across the sample region.